\documentstyle[PASJadd,psbox]{PASJ95}
\newcommand{\lett}{}

\begin{document}
\title{Discovery of a 9-s X-Ray Pulsar, AX~J0049$-$732, \\
in the Small Magellanic Cloud}
\author{Masaru {\sc Ueno}, Jun {\sc Yokogawa}, Kensuke {\sc Imanishi}, 
and Katsuji {\sc Koyama}\thanks{CREST, Japan Science and Technology 
Corporation (JST), 
4-1-8 Honmachi, Kawaguchi, Saitama 332-0012.}\\
{\it Department of Physics, Graduate School of Science, Kyoto
University, Sakyo-ku, Kyoto 606-8502}\\
{\it E-mail(MU): masaru@cr.scphys.kyoto-u.ac.jp}}
\abst{We report on an ASCA discovery of a new X-ray pulsar, 
AX~J0049$-$732, in the Small  Magellanic Cloud (SMC).  The pulse 
shape is sinusoidal with a barycentric period of $9.1320 \pm 
0.0004$~s. The X-ray spectrum was fitted by an absorbed power-law 
model with a photon index of $0.6^{+1.0}_{-0.7}$  and a column 
density of $1.3^{+2.9}_{-1.3} 
\times 10^{22}$~cm$^{-2}$.  An unabsorbed flux at 0.7--10.0 
keV was estimated to be $8 \times 10^{-13}$~erg~cm$^{-2}$~s$^{-1}$ 
corresponding to an X-ray luminosity of $4 \times 
10^{35}$~erg~s$^{-1}$ at an SMC distance of 62~kpc.
}   
\kword{pulsars: individual (AX~J0049$-$732) --- stars: neutron --- 
X-rays: stars}

\maketitle
\thispagestyle{headings}

%section 1
\section{Introduction}

Since the discovery of X-ray pulsation from Cen X-3 (Giacconi et al.\ 1971), 
about 80 X-ray binary pulsars have been discovered  at present 
(e.g., Nagase 1999).  
Most of them are located  
in either the Galactic plane or the satellite galaxies, the Large 
Magellanic Cloud and the Small Magellanic Cloud (SMC).

The SMC, with its reasonable size ($\sim 3^{\circ} \times 
3^{\circ}$), proximity (62~kpc; Laney, Stobie 1994), and  small 
galactic absorption ($\sim 10^{20} $~cm$^{-2}$),
is suitable for population studies of X-ray 
binaries.  At an SMC distance of 62~kpc, an X-ray binary 
pulsar  with a typical 
luminosity of  $\sim 10^{35}$~erg~s$^{-1}$ should be 
observed with a flux of $\sim 10^{-13}$~erg~cm$^{-2}$~s$^{-1}$, 
which is well above the detection limit of the current X-ray 
satellites (ROSAT, ASCA, RXTE, and BeppoSAX). During the last 3 years, 
more than a dozen binary pulsars have been discovered in the SMC (e.g.,\ 
Yokogawa et al.\ 2000). 

Most of the new X-ray binary pulsars either show a transient behavior 
or are associated with a Be star, or both. Hence, X-ray pulsars newly 
discovered in the SMC are naturally classified as Be/X-ray binary 
pulsars (e.g., Yokogawa et al.\ 2000). 
Recent discoveries of transient pulsars have revealed that 
the fraction of Be star binaries in the SMC is much larger than 
that of our Galaxy. 
This suggests an active star-formation history within the past $\sim 
10^{7}$ years (Haberl, Sasaki 2000; Yokogawa et al.\ 2000). 
Encouraged by these fruitful 
outcomes, we have further continued the pulsar search project of the 
SMC. This letter reports on a serendipitous discovery of a new pulsar, 
AX~J0049$-$732, in the SMC with ASCA.

%section 2
\section{Observation and Data Reduction}

ASCA observed an SMC region centered at 
R.A.~= 00$^{\rm h}$47$^{\rm m}$16$^{\rm s}$, 
Decl.~= --73$^{\circ}$08$'$30$''$ 
(J2000) in 1997 November 13--14 
in order to study N~19, a radio supernova remnant in the SMC. 

ASCA (Tanaka et al.\ 1994) has four identical X-ray telescopes
with 
nested thin foil mirrors (XRT; Serlemitsos et al.\ 1995).  The
focal-plane instruments consist of the two Solid-state Imaging
Spectrometers (SISs: Burke et al.\ 1994) and the two Gas Imaging
Spectrometers (GISs: Ohashi et al.\ 1996; Makishima et al.\ 1996).
In this observation, GISs were operated in the normal PH mode, 
providing time 
resolutions of 62.5~ms and 0.5~s in high and medium bit-rates, 
respectively. The SISs were operated in the 2-CCD Faint/Bright 
mode with the a level discrimination of 0.7~keV. 

We screened both the GIS and SIS data using the standard procedure 
by rejecting data obtained in the South Atlantic Anomaly, in low 
cut-off rigidity regions ($<6$~GV for GIS and $<4$~GV for SIS), and 
at an elevation angle lower than $5^{\circ}$. A rise-time 
discrimination was applied to the GIS data to reject particle 
events. Hot and/or flickering pixels were 
rejected from the SIS data. 
The resultant effective exposures after these screenings were $ 
\sim 40$~ks  for GIS and $ \sim 33$~ks for SIS. Since the dark 
current and its pixel-to-pixel fluctuation in the SIS have 
significantly increased after the 5 years of operation in orbit, 
we applied the Residual Dark Distribution correction (T.~Dotani et al.\ 
1997, ASCA News 5, 14) for the faint mode data.

%section 3
\section{Analysis}

%section 3.1
\subsection{Images}

The GIS image in the energy band of 0.7--7.0~keV is 
shown in figure~1.  The bright source at the field center is the radio 
supernova remnant, N~19. Two pulsars, AX~J0049$-$729  and 
AX~J0051$-$733, were serendipitously found, as already reported by 
Yokogawa et al.\ (1999) and Imanishi et al.\ (1999). 
Another pulsar, RX~J0052.1$-$7319 (Lamb et al.\ 1999; Israel et al.\ 
1999) discovered with ROSAT, is also in the field of view and marginally 
detected in this observation. A point source at the south of N~19 
coincides with the position of source No.~1 in Inoue et al.\ (1983) 
(hereafter IKT~1) or 
source No.\ 434 in the ROSAT PSPC catalogue of Haberl et al.\ (2000). 
IKT~1 is the only bright point source which is in the field of SIS. 
To remove systematic 
errors in the coordinates, we shifted the sky coordinates of ASCA so 
that the position of IKT~1 in the ASCA observation 
comes to the position of the ROSAT PSPC catalogue.
 
Since a faint source in the east of N~19 shows a complex spatial 
structure, we made two energy band images of this particular 
region. Figure~2 shows magnified views of the region in both the 
0.7--2.0~keV (soft) and 2.0--7.0~keV (hard) bands, and both with GIS 
and SIS. We can see two sources: 
one in the soft band and the other in the hard band, each separated 
by $2'$. The soft source coincides with the position of 
SNR~0047$-$735 (Haberl et al.\ 2000). Using the SIS hard band 
(2.0--7.0~keV) image, we determined the position of the hard 
source to be R.A.~= 00$^{\rm h}$49$^{\rm m}$13$^{\rm s}$, 
Decl.~= --73$^{\circ}$11$'$42$''$ (J2000) with an error of 
$40''$ radius, hence designated as AX~J0049$-$732 
(Imanishi et al.\ 1998). 

%
% Figure 1
%
\begin{figure}
\hspace*{8mm}\psbox[xsize=0.4\textwidth]{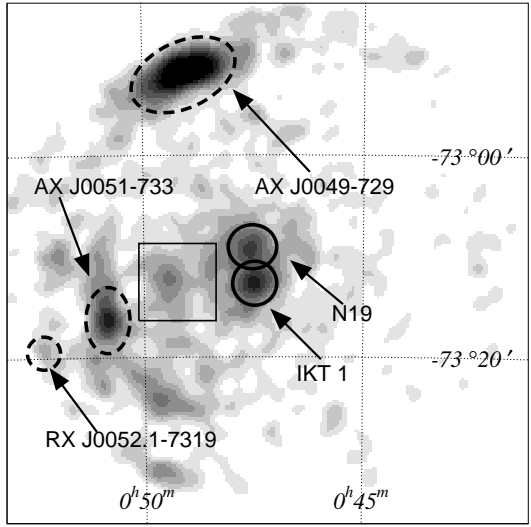}
\caption{ASCA GIS~2 image in the X-ray energy band of 0.7--7.0~keV. 
Three previously known pulsars are designated by broken-line 
ellipses. 
The calibration source region (at the bottom) is discarded.
The magnified images of the region designated by the solid line 
box are shown in figure 2.}
\end{figure}
%
% Figure  2
%
\begin{figure}
\hspace*{8mm}\psbox[xsize=0.45\textwidth]{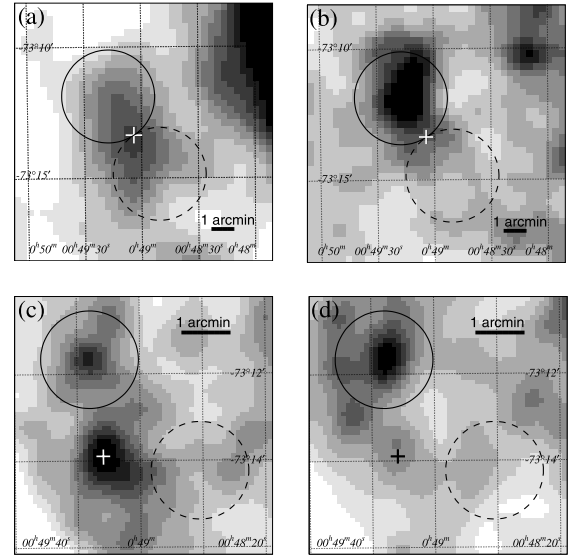}
\caption{Magnified images of the box region in figure 1: 
(a) GIS 0.7--2.0~keV (b) GIS 2.0--7.0~keV (c) SIS 0.7--2.0~keV 
(d) SIS 2.0--7.0~keV. The source and background regions 
for the energy spectrum analysis are shown by the solid-line circles 
and broken-line circles, respectively. SNR~0047$-$735 is designated 
by the crosses.
}
\end{figure}

%section 3.2
\subsection{Energy Spectrum}

The X-ray spectra of AX~J0049$-$732 were made
by accumulating photons in the region of a circle of $2'$ 
radius for GIS and $1'$ radius for SIS, as shown in figure~2. 
In order to remove any possible contamination from SNR~0047$-$735 
located $2'$ away, the background spectra were made 
using data in the regions 
at the same distance ($2'$) from SNR~0047$-$735. These background 
regions are also shown in figure~2. 

% 
% Figure 3
% 
\begin{figure}
\hspace*{8mm}\psbox[xsize=0.45\textwidth]{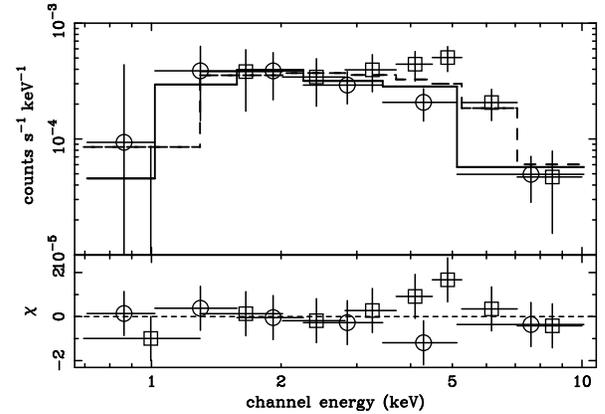}
\caption{Energy spectra of AX~J0049$-$732 obtained from 
the SIS (circles) and GIS (squares). The solid line and the dashed line 
are the best-fit power-law spectra to the SIS and GIS spectra, 
respectively(upper pannel). 
Residuals from the best-fit model(lower pannel).}
\end{figure}

The background-subtracted spectra of GIS and SIS are given in 
figure~3. Since no conspicuous line structure was found, we fitted 
the GIS and SIS spectra to a power-law with soft X-ray absorption 
by the interstellar medium. 
We obtained statistically acceptable fits for both the GIS and SIS
spectra. Since the best-fit parameters were consistent 
with each other, we carried out a simultaneous fit to the combined 
GIS and SIS spectra. The 
fit was acceptable with a $\chi^{2}$ value of 7.76 for 11 
degrees of freedom. The resultant best-fit parameters are 
the photon index $\Gamma = 0.6^{+1.0}_{-0.7}$, 
the normalization $= 3.0^{+ 12.5}_{-2.2}$~photons~s$^{-1}$~keV$^{-1}$ 
cm$^{-2}$ at 1~keV, 
and the absorption column density {\it N}$_{\rm H}$ = $1.3^{+2.9}_{-1.3} 
\times 10^{22}$~cm$^{-2}$. The model spectrum convolved with the 
best-fit parameters   
is shown by the solid and dashed lines in figure~3, respectively, for 
the SIS and GIS. 

The absorbed and unabsorbed X-ray fluxes in the 0.7--10.0~keV band 
were calculated to be  $\sim 6.9 \times 10^{-13}$~erg~cm$^{-2}$ 
s$^{-1}$ and $\sim 7.8 \times 10^{-13}$~erg~cm$^{-2}$~s$^{-1}$, 
respectively. 

%section 3.3
\subsection{Time Variability }

For a timing analysis of AX~J0049$-$732, X-ray photons in the 
1.0--5.1~keV band
were extracted from a circular region of $3'$ radius in the GIS~2 and 
GIS~3 images. After converting to the barycentric arrival time, we 
searched for periodicity using a Fast Fourier Transformation (FFT) 
algorism.  
Figure~4 shows the resultant 
power density spectrum in the $1.0 \times 10^{-5}$--1.0~Hz frequency band. 
A significant peak 
can be clearly seen at 
0.1095~Hz with 99.99\% confidence. Since 
SNR~0047$-$735 is located near AX~J0049$-$732, one may argue 
that the pulsations might originate from a putative neutron star 
in SNR~0047$-$735. To solve this ambiguity we also searched for 
pulsations in the SNR data extracted from a region of the same size 
circle centered on SNR~0047$-$735. However, no significant pulsation was 
found in the power spectrum of the SNR~0047$-$735 data. Therefore, 
it is concluded that the 0.1095~Hz pulsations are not 
attributed to SNR~0047$-$735, but to AX~J0049$-$732. 

To determine the pulse period more precisely, we performed an epoch
folding near the trial period found with the FFT; 
barycentric pulse period of 9.1320 $\pm$ 0.0004~s was obtained.
The pulse profile folded with this period is shown in figure~5.
The profile is nearly sinusoidal  with a pulse fraction, defined as 
(pulsed flux)/(total flux), of $\sim 56$\%.

We examined the aperiodic intensity variation in the 0.7--7.0~keV 
light curve of the whole observation period. Neither a significant flux 
variation, nor any burst-like activity was found on the time scale from 
seconds to hours.

%
% Figure 4
%
\begin{figure}
\psbox[xsize=0.45\textwidth]{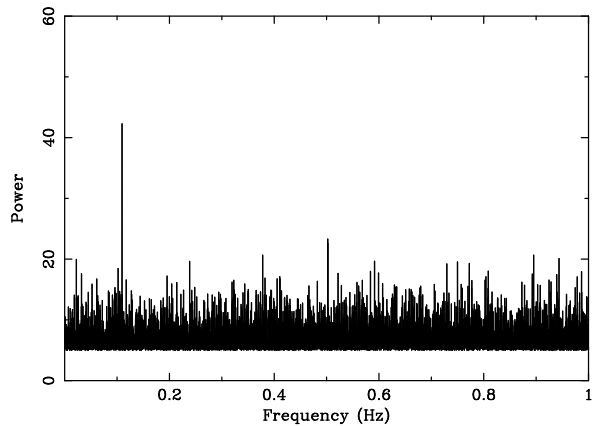}
\caption{Power-density spectrum of AX~J0049$-$732 derived from the GIS data in 
the 1.0--5.1~keV band. The power is normalized to be 2 for random 
fluctuations. 
Data points smaller than 5 are not plotted.} 
\end{figure}

%
% Figure 5
%
\begin{figure}
\psbox[xsize=0.45\textwidth]{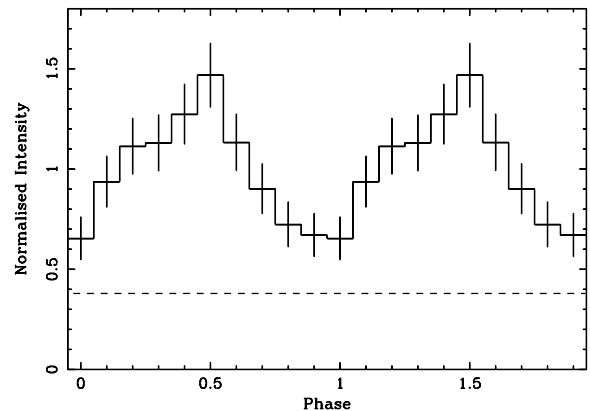}
\caption{ Light-curve folded by the best-fit period for the GIS 
data in the 1.0--5.1~keV band. Two cycles are shown. The dashed line 
indicates the background level.}
\end{figure}

\section{Discussion}
At the distance of the SMC, the intrinsic luminosity of AX~J0049$-$732
is estimated to be $4 \times 10^{35}$~erg~s$^{-1}$, which is below 
the range of bright X-ray pulsars. Thus, AX~J0049$-$732 would not
be a member of luminous X-ray pulsars powered by Roche-lobe 
over-flow from a companion star. Although the observed luminosity 
suggests that the pulsar might be a wind-fed supergiant system, 
the short spin period of 9~s does not favor this class, which typically 
shows a spin period of 100--1000~s.

A more likely scenario for  AX~J0049$-$732 is either a Be star X-ray 
binary system, or an anomalous X-ray pulsar (e.g., Bildsten et al.\ 1997; 
Mereghetti, Stella 1995). Direct information to distinguish 
these two possibilities can be obtained by measuring the pulse period 
derivative and its orbital modulation. Limited photon statistics and 
observation coverage, however, do not allow us to derive any constraints 
on the pulse period derivative and the orbital motion. Nevertheless, 
the hard X-ray spectrum with a photon index of $\sim$ 0.6 favors a Be star 
X-ray binary system for the pulsar. The existence of a Be star counterpart, 
if confirmed, would be strong evidence for this scenario.

Two sources, No.\ 427 and No.\ 430, in the ROSAT PSPC catalogue 
of Haberl et al.\ (2000) are possible counterparts of AX~J0049$-$732 
located at distances of $1'\hspace{-2.8pt}.43$ and $0'\hspace{-2.8pt}.15$, 
respectively. Filipovi\'{c} et al.\ (2000) searched for optical counterparts 
of these ROSAT sources, and found an emission line object, possibly a Be 
star, at the position of source No.\ 427, but found no counterpart for 
source No.\ 430. 
Hence, they suggest that source No.\ 427 is more likely to be 
a counterpart of AX~J0049$-$732. This possibility, 
however, may be rejected, because the angular separation of these 
sources of $1'\hspace{-2.8pt}.43$ is significantly larger than the ASCA error radius.
We rather propose that No.\ 430 is a more likely counterpart.
The flux of source No.\ 430 was roughly the same as that of AX~J0049$-$732, 
although systematic errors in the flux detemination make it 
ambiguous to compare the fluxes by different instruments. 
We searched for an optical object in the error circle of AX~J0049$-$732 
using the SIMBAD database, but found no optical counterpart.
Although a Be/X-ray binary pulsar is likely, we thus need further 
X-ray observations and a search of an optical counterpart to establish 
the true nature of this new pulsar.  

AX~J0049$-$732 is the fourth X-ray pulsar discovered in the 
small area of the SMC South (within $30'$ radius from N~19). The 
three pulsars previously reported are all identified to Be star 
X-ray binary systems (Haberl, Sasaki 2000; Israel et al.\ 1999), 
and the fourth pulsar, AX~J0049$-$732, can 
probably be classified into the same class. The high density of 
Be/X-ray binaries of this region in the SMC suggests an active 
star-formation history of this 
area within the past $\sim 10^{7}$ yrs which is the typical age 
of Be star X-ray binary pulsars. We note that the 
four pulsars are all located along the edge of the 
H$_{\rm I}$ supergiant shell (Stanimirovi\'{c} et al.\ 1999), whose age 
is estimated to be about $10^{7}$ years. The supergiant shell may be 
other evidence for the star-formation history. 

\par
\vspace{1pc}\par
M.U.\ thanks Prof.\ Kamae for the aid during his stay in the University 
of Tokyo where this paper was prepared. We thank all members of 
the ASCA team. 
J.Y.\ is supported by JSPS Research Fellowship for Young Scientists.

\section*{References}
\small

\re
 Bildsten L., Chakrabarty D., Chiu J., Finger M.H., Koh D.T., 
 Nelson R.W., Prince T.A., Rubin B.C.\ et al.\ 1997, ApJS 113, 367
\re
 Burke B.E., Mountain R.W., Daniels P.J., Dolat V.S., 
 Cooper M.J.\ 1994, IEEE Trans.\ Nucl.\ Sci. 41, 375
\re
 Giacconi R., Gursky H., Kellogg E., Schreier E., Tananbaum H.\ 1971,
 ApJ 167, L67
\re 
 Filipovi\'{c} M.D., Pietsch W., Haberl F.\ 2000, A\&A 361, 823
\re
 Haberl F., Filipovi\'{c} M.D., Pietsch W., Kahabka P.\ 2000, A\&AS 
 142, 41
\re
 Haberl F., Sasaki M.\ 2000, A\&A 359, 573
\re
 Imanishi K., Yokogawa J., Koyama K.\ 1998, IAU Circ. 7040 
\re
 Imanishi K., Yokogawa J., Tsujimoto M., Koyama K.\ 1999, PASJ 51, 
L15
\re 
Inoue H., Koyama K., Tanaka Y.\ 1983, Supernova 
  Remnants and Their X-ray Emission,  in IAU Symp.\ 101, ed J.\ Danziger, 
P.\ Gorenstein (Dorderecht, Reidel) p535
\re
  Israel, G.L., Stella, L., Covino, S., Campana, S., Mereghetti, S.\ 
  1999, IAU Circ. 7101
\re
 Lamb R.C., Prince T.A., Macomb D.J., Finger M.H.\ 1999, IAU Circ. 
7081
\re
 Laney C.D., Stobie R.S.\ 1994, MNRAS 266, 441
\re
 Makishima K., Tashiro M., Ebisawa K., Ezawa H., Fukazawa Y., 
Gunji S., Hirayama M., Idesawa E.\ et al.\ 1996, PASJ 48, 171
\re
 Mereghtti S., Stella L.\ 1995, ApJ 442, L17
\re
 Nagase F.\ 1999, Proc.\ of X-ray Astronomy 1999, ISAS Research 
 Note 694
\re
 Ohashi T., Ebisawa K., Fukazawa Y., Hiyoshi K., 
 Horii M., Ikebe Y., Ikeda H., Inoue H.\ 
 et al.\ 1996, PASJ 48, 157
\re
 Serlemitsos P.J., Jalota L., Soong Y., Kunieda H., Tawara Y., 
 Tsusaka Y., Suzuki H., Sakima Y.\ 
 et al.\ 1995, PASJ  47, 105
\re
 Stanimirovi\'{c} S., Staveley-Smith L., Dickey J.M., Sault R.J., 
Snowden S.L.\ 1999, MNRAS 302, 417
\re
 Tanaka Y., Inoue H., Holt S.S.\ 1994, PASJ 46, L37
\re
 Yokogawa J., Imanishi K., Tsujimoto M., Kohno M., Koyama K.\ 1999, 
PASJ 51, 547
\re 
 Yokogawa J., Imanishi K., Tsujimoto M., Nishiuchi M., Koyama K., 
 Nagase F., Corbet R.H.D.\ 2000, ApJS 128, 491

\end{document}